\documentclass[reprint,
groupedaddress,
 amsmath,amssymb,
 aps,
prx,
onecolumn]{revtex4-2}

\usepackage{graphicx}
\usepackage{comment}
\usepackage{epsfig}
\usepackage{epstopdf}
\usepackage{color}
\usepackage{gensymb}
\usepackage{dcolumn}
\usepackage{bm}
\usepackage{amsmath,amssymb}
\usepackage[colorinlistoftodos]{todonotes}

\def\.{\cdot}

\def\_#1{{\bf #1\mit}}
\def\=#1{\overline{\overline #1}}

\def\##1{{\bf #1\mit}}
\def\_#1{{\bf #1\mit}}
\def\=#1{\overline{\overline #1}}

\def\e{\begin{equation}}
\def\f{\end{equation}}

\graphicspath{{figures/}}
\begin{document}

\title{Electromagnetic Time Interfaces in Wire Media: Innovations for Subwavelength Imaging}

\author{Constantin~Simovski$^1$}
\email{konstantin.simovski@aalto.fi}  
\author{Mohammad~Sajjad~Mirmoosa$^2$} 
\author{Sergei~Tretyakov$^1$}

\affiliation{$^1$Department of Electronics and Nanoengineering, Aalto University, P.O.~Box~15500, FI-00076 Aalto, Finland\\$^2$Department of Physics and Mathematics, University of Eastern Finland, P.O.~Box 111, FI-80101 Joensuu, Finland} 



\begin{abstract}
In this work, we theoretically study temporal interfaces between media with strong spatial dispersion and dielectrics. In particular, we consider a temporal discontinuity that transforms a wire medium sample -- a metamaterial with resonant spatial dispersion -- into a uniaxial dielectric. We show that this transition results in a transformation of the deeply subwavelength spatial spectrum of TEM waves propagating in the wire medium at a certain frequency into a spectrum of plane waves at new frequencies 
that are all higher than the initial one. The waves at different frequencies propagate in different directions. Their complex amplitudes and propagation directions are uniquely related to the amplitudes of the spatial harmonics of the fields which existed before the transition. We explain how to implement this transition. The revealed effect may result in a promising method of subwavelength imaging. 
\end{abstract} 


\maketitle

\section{Introduction} 

In recent years, there has been a surge of interest in analyzing electromagnetic time-varying systems, e.g.~\cite{galiffi2022photonics,ptitcyn2023tutorial}, particularly focusing on temporal discontinuities~\cite{mostafa2024temporal}. This research domain considers sudden changes at specific moments of time in effective parameters that characterize the system response. In the realm of electromagnetic media, these abrupt jumps, such as in the permittivity tensor, give rise to time interfaces between different media. 
{Generation of reflected waves, frequency conversion, breaking the conservation of power~\cite{morgenthaler1958velocity,mendoncca2002time,Agrawal2014RTC}, and photon-pair creation~\cite{mendoncca2000quantum,mirmoosa2023quantum} are these interfaces' four fundamental features that have been contemplated theoretically~\cite{morgenthaler1958velocity,Wilks1988TH,mendoncca2002time,Agrawal2014RTC,mendoncca2000quantum,mirmoosa2023quantum} and studied experimentally~\cite{Yugami2002EXPER,nishida2012EXP,water,Alu_TL}. Considering such interfaces in complex media results in uncovering a multitude of functionalities including polarization conversion~\cite{xu2021complete}, polarization-dependent analog computing~\cite{RizzaCastaldiGaldi23}, polarization splitting~\cite{mostafa2023spin}, polarization rotation~\cite{mirmoosa2024time}, inverse prism~\cite{akbarzadeh2018inverse}, anti-reflection temporal coatings~\cite{ramaccia2020light,pacheco2020anti}, temporal aiming~\cite{pacheco2020aiming}, wave freezing and thawing~\cite{wang2023controlling}, wiggler-mode production~\cite{jiang1975wave, kalluri1988reflection, Wilks1988TH}, a transformation of surface waves into free-space radiation~\cite{Grap19SPP,wang2023controlling}, direction-dependent wave manipulation~\cite{mirmoosa2024time}, and so forth.}

Despite the extensive exploration in this field, investigations into time interfaces of media with strong spatial dispersion remain conspicuously absent. One specific example of those media is a metamaterial of thin parallel wires called wire medium (WM) {whose electromagnetic properties and prospective applications were thoroughly reviewed in~\cite{simovski2012wire}. Among these applications, the most relevant for the present paper is the radio-frequency, microwave, THz and infrared endoscopy which grants subwavelength resolution of imaging devices. The subwavelength resolution of corresponding WM endoscopes was experimentally demonstrated in papers \cite{Slobo,belov2005canalization,belov2008transmission,tuniz2013metamaterial,casse2010super} and several further works. After publication of the overview~\cite{simovski2012wire} some new applications of WM were found, such as radiative cooling studied in works \cite{hypersim,Kos1,Kos2}.}  

{Here, we study the instantaneous transition of a  wire medium into a uniaxial dielectric. This discontinuity transforms the strongly spatially dispersive WM of continuous wires into an artificial uniaxial dielectric with negligible spatial dispersion.  At this stage, we analyze a temporal discontinuity in an infinitely extended WM sample. We  deduce the amplitudes of the time-refracted and time-reflected waves at new frequencies and analytically show that at such temporal interfaces, conversion of subwavelength field distributions into propagating waves takes place. Due to this effect, the spatial variations of the field distribution that exist before the parameter jump are  ``encoded'' in both the frequency spectrum and directivity pattern of emitted radiation. Based on the theoretical results, we show that an implementation of such  temporal discontinuities grants a promising method of subwavelength imaging at THz frequencies and propose an alternative approach to subwavelength imaging using switchable wire-medium endoscopes. In this scenario, there is no need for an electrical connection of the WM sensor with the image acquisition device. The image can be restored at a distance by a spectroscopic measurement of the angular pattern produced by the WM sensor after the temporal jump. 
}


\section{Theory}

\begin{figure*}[t]
\centering
\includegraphics[width=14cm]{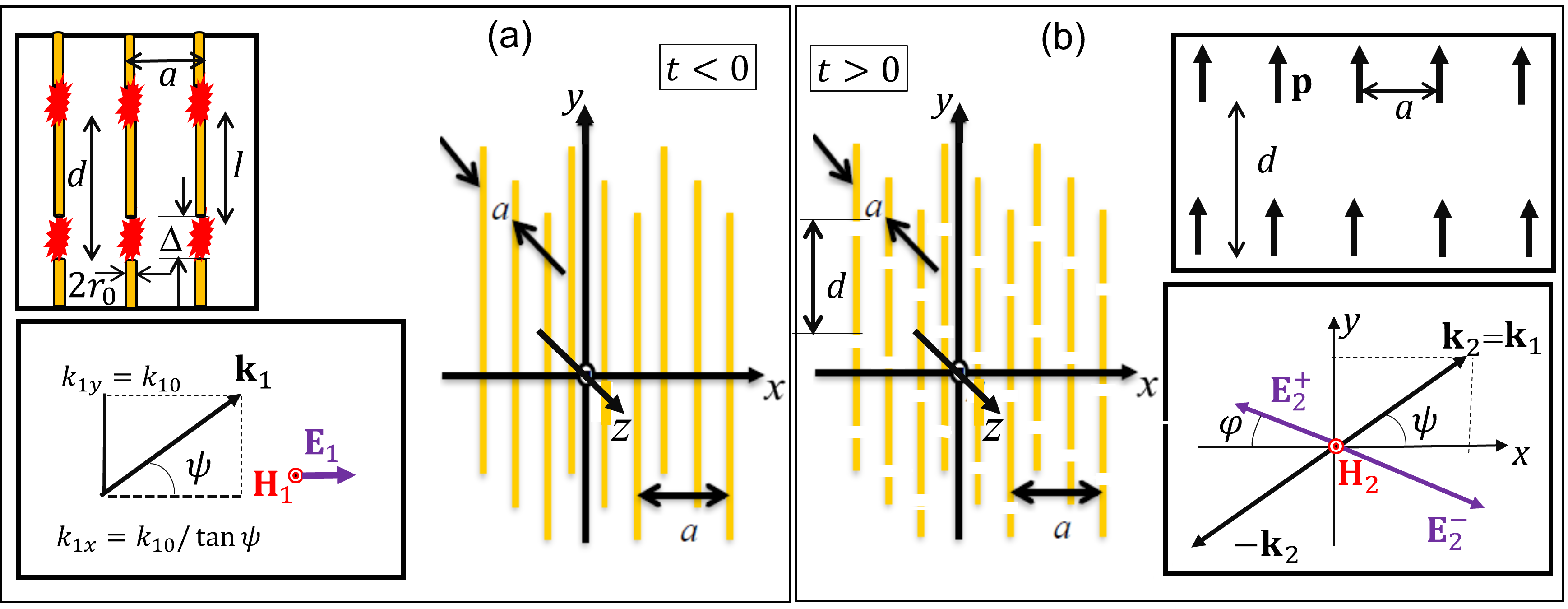}
\caption{(a) Wire medium of solid wires in which a non-uniform TEM wave with the {nonzero} spatial frequency $k_{1x}$ propagates along $y$. The wavevector $\_k_1$ describes the phase velocity tilted to the wires (bottom inset). Continuous wires can be obtained from split wires using, for example, microplasma gas discharge switches (top inset).  (b) Wire medium of split wires in which the time-refracted and the time-reflected TM-polarized waves arise. This medium is a uniaxial lattice of electric dipoles (top inset). On the bottom inset the vectors of the time-refracted and time-reflected waves arising in this medium are shown.}
\label{fig1}
\end{figure*}

In a wire medium (WM) of continuous perfectly conducting wires, eigenmodes at frequencies below the plasma frequency $\omega_p$ are transverse electromagnetic (TEM) waves propagating along the wires without dispersion. Both phase and group velocities along the wires are equal to $c/n$, where $c=1/\sqrt{\varepsilon_0\mu_0}$ is the speed of light in vacuum, and $n=\sqrt{\varepsilon_h}$ is the refractive index of the host medium \cite{belov2003strong}. Meanwhile, the spatial frequencies, i.e., the transverse wavenumbers of these waves occupy an ultra-broad band from zero to $\pi/a$, where $a$ is the period of the WM. The waves with larger transverse wavenumber cannot propagate along the wires at the frequency $\omega_1$ which is smaller than the effective plasma frequency $\omega_p$ \cite{maslovski2002wire}. {It means that the wave in the WM is transverse with respect to the group velocity and Poynting vector. Meanwhile, the wavevector $\_k_1$ comprises both $x$- and $y$-components and may be tilted with respect to the wires under the nonzero angle $\pi/2-\psi$, as shown in Fig. \ref{fig1}.}

In accordance with \cite{belov2003strong}, a lossless WM is described by the spatially dispersive permittivity tensor with the axial component
\e
\varepsilon^{(1)}_{yy}=\varepsilon_{h}\left(1-{k_p^2\over k_{10}^2-k_{1y}^2}\right).
\label{eps}\f
In Eq.~\eqref{eps}, $\varepsilon_h=n^2$ is the relative permittivity of the host medium, $k_p\equiv n\omega_p/c$ denotes the effective plasma wavenumber of the WM given by $k_{p}^2={2\pi \varepsilon_{h}/[a^2\log{a^2/4r_0(a-r_0)}}]$ \cite{maslovski2002wire} (where $r_0$ is the wire cross-section radius, and $a$ is the period), and $k_{10}=n\omega_1/c$ represents the wavenumber of a wave that would propagate in the host medium without wires at the frequency $\omega_{1}$. The spatial dispersion is manifested in Eq.~\eqref{eps}  by the dependence of the axial permittivity on the wavevector component $\_y_0k_{1y}$ {(here and below $\_y_0$, as well as $\_x_0$ and $\_z_0$, denote the unit vectors of the Cartesian coordinate axes)}. If the wires are thin enough, i.e.~$r_0\le 0.1a$ (see, e.g.,~Ref.~\cite{maslovski2002wire}), the transverse component $\varepsilon^{(1)}_{xx}=\varepsilon^{(1)}_{zz}$ of the permittivity tensor is practically not affected by the wires and approximately equals $\varepsilon_h$. Here, we orient the axis $z$ so that there is no field variation along it and consider the case when the magnetic field $\_H_1$ is polarized along $z$. 


{Expression \eqref{eps} allows one to see that} the low-frequency ($\omega_1<\omega_p$) eigenmodes of the WM ($|k_{1x}|<1/a$) are non-uniform plane waves that have the same group velocity and differ from one another only by their spatial frequencies $k_{1x}$. In Fig.~\ref{fig1}(a), the spatial frequency is shown by a dashed arrow to stress that this value describes only the phase distribution across the energy propagation direction. Since the phase front of the TEM wave is tilted by the angle $\pi/2-\psi$ to the plane $(x-z)$, the vector $\_k_1$ corresponds to the phase velocity $\omega_1/k_1=c\sin\psi/n$.  The electric and magnetic field vectors are, respectively, $\_E_1=E_1\_x_0$, $\_H_1=H_1\_z_0=E_1\_z_0/\eta$, where $\eta=\sqrt{\mu_0/\varepsilon_0\varepsilon_h}$ is the wave impedance of the host medium \cite{belov2003strong}. Further, for simplicity of writing, we assume that the host medium of our wires is air ($\varepsilon_h=1$).

Let at the moment $t=0$, simultaneous breaks occur in every wire at a sufficiently small period 
$d\ll\lambda_1$, where $\lambda_1=2\pi c /\omega_1$ (recall that this wavelength is much larger than $a$ since 
$\omega_1\ll \omega_p$). Consequently, the WM of solid wires (WM~1) transforms into a WM of split wires (WM~2). In this transition, waves at a new frequency $\omega_2$ arise in accordance with the general theory of temporal discontinuities in materials~\cite{morgenthaler1958velocity}. Assume that the splits are of substantial length ($\Delta\gg r_0$, where $\Delta$ is the split width and $r_0$ is the wire radius) and that the period $d$ remains subwavelength also at $\omega_2$. Hence, even in the case $\Delta\ll d$, WM~2 will be a regular metamaterial of short metallic rods of length $l=d-\Delta$, and its permittivity can be found as~\cite{collin1990field}: 
\e
\varepsilon^{(2)}_{yy}=1+{\alpha/V\over 1-\alpha C},
\label{exp}\f
where $V=da^2$ is the lattice unit cell volume, $C=1.2/(\pi d^3)-(8\pi/d^3)K_0(2\pi a/ d)$ is the lattice interaction factor ($K_0$ is the McDonald function), and $\alpha$ is the polarizability of the rod of length $l$ that is defined in \cite{collin1990field} as $\alpha\equiv p/\varepsilon_0E_{\rm ly}$. Here, $p=p_y$ is the rod dipole moment, $E_{\rm ly}$ is the $y$-component of the local electric field at the rod center. If $a=d$, the formula for $C$ gives $C=1/3$, and the expression for $\varepsilon^{(2)}_{yy}$ transforms into the well-known Lorenz-Lorentz formula. In Ref.~\cite{collin1990field}, $\alpha$ is given only for spheroidal metal particles.  Formula (5.17) from~\cite{tretyakov2003analytical} gives for a short thin rod $\alpha=4l^2/(3j\omega_2 \varepsilon_0 Z_{sd})$, where $Z_{sd}$ is the complex input impedance of a short dipole~\cite{balanis2016antenna}. The real part of this impedance corresponds to the imaginary part of $\alpha$, which describes the scattering by the rod. This imaginary part must be omitted in the expression for effective permittivity. An accurate model of a uniaxial lattice \cite{belov2005homogenization} shows that the scattering part of the rod polarizability cancels out with the imaginary part of the interaction constant $C$ (this imaginary part is responsible for the field retardation on the scale of a unit cell). Therefore, the effective permittivity of a lossless lattice is obviously real and in its expression we should replace the complex polarizability $\alpha$ by its real part $\alpha'$ substituting  $Z_{sd}=1/j\omega_2 C_{sd}$. Here, $C_{sd}=\pi\varepsilon_0l/2\log{(l/r_0)}$ is the capacitance of the short dipole (see e.g. in \cite{balanis2016antenna}). In the expression for $\alpha'$, the frequency $\omega_2$ cancels out, and the expression \eqref{exp} takes the form 
\e
\varepsilon^{(2)}_{yy}\approx 1+{\alpha'\over V(1-\alpha' C)},
\label{eps2}\f 
where $\alpha'=l^3/[6\log{(l/r_0)}]$. Equation~\eqref{eps2} means that our WM~2 is a dispersion-free uniaxial dielectric. 

Let us see what happens if the WM of continuous wires transits to such a dielectric. In accordance with Ref.~\cite{morgenthaler1958velocity}, at $t>0$, both time refraction and time reflection must occur: Two waves with electric field amplitudes $\_E_2^{\pm}$ and magnetic ones $\_H_2^{\pm}$ oscillating at a new frequency $\omega_2$ will propagate in opposite directions, as it is shown in 
Fig.~\ref{fig1}(b). The time-refracted wave inherits the previous wavevector: $\_k_{2}=\_k_{1}$. Meanwhile, the tilt angle $\phi$ of the electric field vector $\_E_2^{+}$ of the time-refracted wave is different from $\pi/2-\psi$ since this wave is not TEM. The time-reflected wave propagates oppositely to the time-refracted one and also has the transverse magnetic (TM) polarization, as depicted in Fig.~\ref{fig1}(b). To find the amplitudes of both these waves and the new frequency $\omega_2$, we may use the time boundary conditions for vectors $\_D$ and $\_B$ \cite{morgenthaler1958velocity} that can be written as follows: 
\e
D_{1x}=\varepsilon_0 E_1=D_{2x}=\varepsilon_0 (E_{2}^--E_{2}^+)\cos\phi,
\label{Morg1}\f
\e
B_{1}=\mu_0 {E_1\over\eta}=B_{2}=\mu_0{(E_{2}^++E_{2}^-)\cos\phi\over Z^{TM}},
\label{Morg2}\f
where the uniaxial impedance $Z^{TM}$ relates the transverse components of the electric and magnetic fields ($E_{2x}^{\pm}= \pm E_{2}^{\pm}\sin\phi=\pm Z^{TM}H_{2}^{\pm}$). The time boundary condition for the axial component of the electric displacement ($D_{1y}=D_{2y}$), though holds, is useless because $\varepsilon_{yy}^{(1)}$ is infinite whereas the electric field component $E_{1y}$ is null. 
{Indeed, initially, in WM~1 the electric displacement is fully determined by the currents flowing in the wires. However, we do not know these currents. To measure them is hardly realistic. The theory of wire media does not establish a unique relation between them and the electromagnetic field of the TEM mode.} 


\begin{figure*}[t]
\centering
\includegraphics[width=16cm]{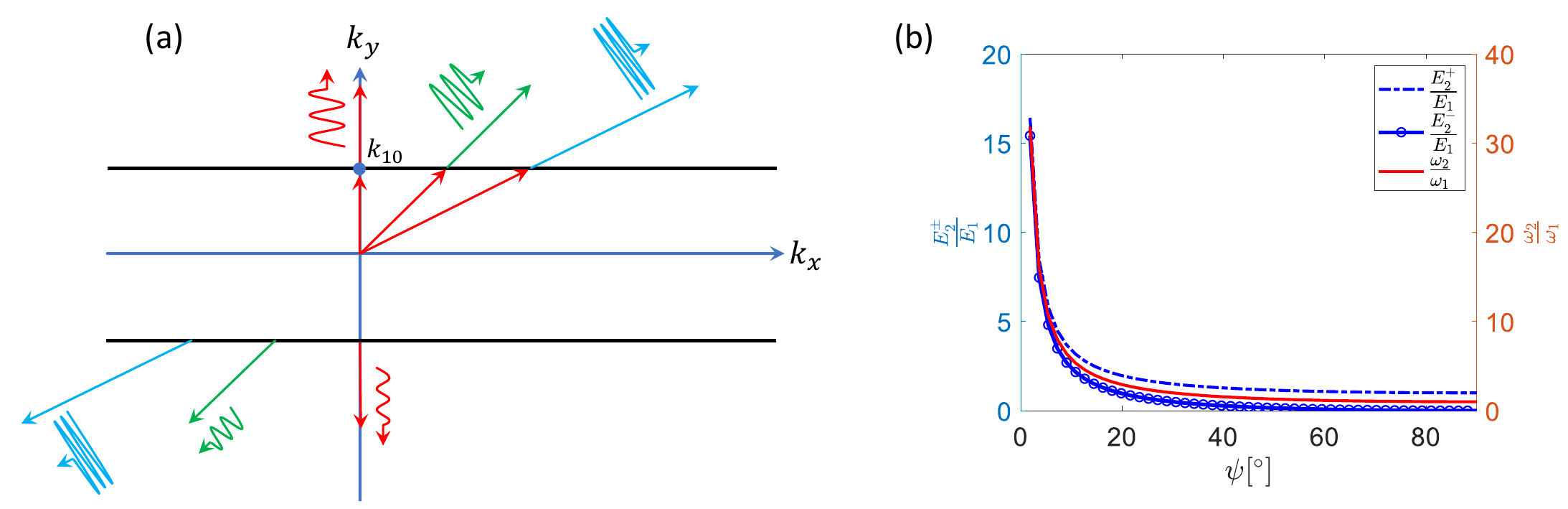}
\caption{(a) The isofrequency contour associated with the wire medium made of perfectly conducting solid wires. The red arrows show different wavevectors for waves propagating in this medium  at a fixed angular frequency. The red, green, and blue signals show the transformed waves traveling at different angular frequencies and amplitudes in the wire medium made of split wires. (b) The amplitudes of the time-reflected (blue curve with markers) and time-refracted waves (dot-dash blue curve) and the ratio $\omega_2/\omega_1$ (red curve) versus the angle $\psi$.}
\label{fig2}
\end{figure*}


From Maxwell's equations for $t>0$ with the substitution $D_{2y}/\varepsilon^{(2)}_{yy}=\varepsilon_0(E_{2}^- - E_{2}^+)\sin\phi$, we obtain $\cot{\phi} =\varepsilon^{(2)}_{yy}\tan \psi$, $Z^{TM}=\eta(1-{k_{2x}^2/ k_{20}^2\varepsilon^{(2)}_{yy}})^{1/2}$, and $k_{2x}E_{2y}^{\pm}-k_{2y}E_{2x}^{\pm}=\eta k_{20}H_{2}^{\pm}$, 
where it is denoted $k_{20}\equiv \omega_2/c$.  
Using the continuity of the axial and transverse wavenumbers $k_{2y}=k_{1y}=k_{10}$, $k_{2x}=k_{1x}\equiv k_{10}\cot{\psi}$, we can write the dispersion equation of the disconnected WM in the form 
$k_{2x}^2/\varepsilon^{(2)}_{yy}+{k_{2y}^2}=k_{10}^2(1+\cot^2\psi/\varepsilon^{(2)}_{yy})= k_{20}^2$, which results in the new frequency (recall that $k_{10}=\omega_1/c$ and $k_{20}=\omega_2/c$) 
\e
\omega_2=\omega_1/\cos\phi=\omega_1\sqrt{1+{\cot^2\psi/\varepsilon^{(2)}_{yy}}}.  
\label{omega2}
\f
In Eq.~\eqref{omega2} for $\varepsilon^{(2)}_{yy}$, we can employ the relation expressed in Eq.~\eqref{eps2}. The wave impedance in WM~2 takes the form 
$Z^{TM}=\eta/\sqrt{1+\cot^2\psi/\varepsilon^{(2)}_{yy}}$. By using this expression with relations \eqref{Morg1} and \eqref{Morg2}, we obtain after some algebra the amplitudes of both time-refracted and time-reflected waves: 
\e
E_2^{\pm}=\pm{E_1\over 2}\left(1\pm{\omega_2\over \omega_{1}}\right).
\label{final}\f
In the case of ultimately high spatial frequencies {$k_{1x}\gg k_{10}$}, i.e., when $\cot\psi$ approaches  $(k_{10}a)^{-1}$, the initial wave is maximally affected by the transformation of WM~1 into WM~2. In this case, both time-refracted and time-reflected waves propagate almost {along $x$}, and  \eqref{final} gives their approximate equivalence: $E_{2}^-\approx E_{2}^+\gg E_1$. The new frequency $\omega_{2}$ is very high and nearly equals $\omega_{1}[\varepsilon^{(2)}_{yy}]^{-1/2}(k_1a)^{-1}$. 
If the spatial frequency is zero, i.e., the initial wave is a uniform plane wave propagating along the wires, it cannot be affected by this temporal discontinuity. Equations~\eqref{omega2} and \eqref{final} give the same amplitude $E_1$ and frequency $\omega_1$ for the time-refracted wave, whereas no time-reflected wave is excited. {All of the theoretical expectations outlined above are illustrated in a schematic representation in Fig.~\ref{fig2}(a) and supported by numerical calculations presented in Fig.~\ref{fig2}(b). For the latter, we employed typical parameters of a wire medium operating in the millimeter wave range.}


\section{Time jumps in inductively loaded wire media}

Now let us discuss how to transform a continuous WM into a cut-wire lattice in practice. Such transformation between arrays of continuous wires and arrays of disconnected pieces of wires can be done using some kind of switches periodically inserted into the wires. To account for finite impedances of the switches, we use 
{the model of a periodically loaded WM that was developed in \cite{maslovski2009nonlocal}. If the distance between insertions is electrically small, the periodical loading of wires with the lumped impedance $Z_{\rm load}$ and the subwavelength period $d$ can be considered as an effectively uniform impedance loading. The impedance per unit length 
$Z_w=Z_{\rm load}/d$ should be added to the wire resistance p.u.l. Then, the axial permittivity of WM~2 becomes \cite{maslovski2009nonlocal}:
\e
\varepsilon^{(2)}_{yy}=1-{k_p^2\over k_{10}^2-j\xi k_{10} - k_{1y}^2},
\label{eps21}\f
where (in the case of perfectly electrically conducting rods) the factor $\xi$ arises due to the loads: 
\e
\xi={\sqrt{\mu_0\varepsilon_0} Z_w\over L_{WM}}, \quad L_{WM}={\mu_0\over 2\pi} \log\left[{a^2\over 4r_0(a-r_0)}\right]. 
\label{xi}\f
Here, $L_{WM}$ is the collective inductance p.u.l. of the unloaded WM (that is, WM~1).  
At a first glance, for splitting wires in WM~1 one may use such electronic switches as pin-diodes, varactors, or MEMS periodically inserted in small gaps in the wires. In the closed state they usually have some parasitic capacitive impedance $Z_{\rm load}$ which can be
negligibly small compared to $\omega_1L_{WM}$, and in the open state they have a large $Z_{\rm load}$.  However, our analysis has shown that the open-state impedance is not large enough. In the switched-off state, all these switches have a non-negligible capacitance, so that the ends of the cut pieces of the wires are coupled much stronger than the via free-space gaps.} 

{Therefore we propose another approach. Let our WM~1 be loaded by inductive loads of finite length. If we remove these connecting loads,  the WM~2 will be the needed array of dipole scatterers, since there will be enough large gaps between wire sections. The inductance of these connectors will add to the effective inductance of WM~1, but qualitatively will change nothing.}  
A  suitable tool that can enable this approach is gas discharge. A usual discharge such as the static arc is difficult to use in practice as it requires high  voltages applied along the wires. However, an electromagnetically assisted discharge across a tiny gap which is utilized in some nanotechnologies \cite{mariotti2010microplasmas} appears to be suitable. If the wire has the thickness of a few microns and the gap $\Delta$ has the width in the interval $\Delta=10-100\, \mu$m, glow microplasma is formed in the gap under practically realistic voltages (dozens of V) and realistic microwave pumping (few mW per one gap) \cite{mariotti2010microplasmas}. In this scenario, every wire of WM~1 should be an array of metal rods whose ends are connected by microplasma channels, as it is sketched in the inset of Fig.~\ref{fig1}(a).  

The impedance properties of the microplasma in such channels were studied in \cite{chen2002impedance} for sub-GHz frequencies. In \cite{gautam2021high}, the impedance of a microplasma capacitor with a large (1~mm) gap between the electrodes was studied in the range $0.3-20$~GHz. Both these studies showed that the channel resistance per unit length drops versus frequency, and at microwaves it is negligibly small compared to the channel inductive impedance. Since the last one grows versus frequency linearly up to approximately 20~GHz, one may assume that the channel inductance weakly depends on the frequency also in the THz range, at least in its low-frequency part. To our estimations, the microplasma discharge of length 20~$\mu$m  has the inductive impedance of the order of  $|Z_{\rm load}|=\omega L_{\rm channel}\sim 2$~Ohm at 100~GHz. 

In accordance with \cite{maslovski2009nonlocal}, we have $Z_{\rm load}=j\omega_1 L_{\rm load}$ and the impedance per unit length 
$Z_w=j\omega_1 L_{\rm load}/d$ arises in WM~1, whose axial permittivity becomes equal to
\e
\varepsilon^{(1)}_{yy}=1-{k_p^2\over k_{10}^2-j\xi_1 k_{10} - k_{1y}^2},
\label{eps22}\f
where the factor $\xi_1$ is now equal to: 
\e
\xi_1=j{k_{10} L_{\rm load}\over d L_{WM}}. 
\label{xi1}\f
Substitution of \eqref{xi1} into \eqref{eps22} transforms this relation to
\e
\varepsilon^{(1)}_{yy}=1-{k_p^2\over k_{10}^2(1+\beta) - k_{1y}^2}, \quad \beta\equiv {L_{\rm load}\over d L_{WM}}.
\label{eps23}\f
In this model, the eigenmodes of WM~1 are quasi-TEM waves with the 
wavevector $\_y_0k_{1y}$, whose absolute value is equal to $k_{1y}=k_1\sqrt{1+\beta}$. It makes the derivations more cumbersome, but the results \eqref{omega2} and \eqref{final} remain valid with a 
minor change: the multiplicative factor $\sqrt{1+\beta}$ arises in their right-hand sides.  
This constant factor changes qualitatively nothing in the curves depicted in Fig. \ref{fig2}(b).


\section{Application of the revealed effect for subwavelength imaging}

Let us discuss an interesting implication of {the revealed phenomenon. It is known that the conversion of subwavelength field distributions into propagating modes that carry the information on the subwavelength features of the input package of evanescent spatial harrmonics can enable hyperlensing \cite{Narimanov,Shalaev,hyperreview,hypersim}. Realizations of hyperlenses are based on the use of complex materials with hyperbolic dispersion or tapered endoscopes -- wire-medium filled cables with gradually increasing distances between the wires. These endoscopes are used for THz subwavelength imaging, and their resolution and maximal length are severely restricted by losses. In this study, we show a way to a much more efficient conversion of the evanescent waves into propagating modes using time jumps of spatially uniform wire media.}

Indeed, in accordance with Eq.~\eqref{omega2}, the  waves propagating in WM~1 with large spatial frequencies corresponding to the condition $k^2_{1x}\gg k_{10}^2\varepsilon^{(2)}_{yy}$  transform into plane waves propagating in WM~2 with the frequencies $\omega_2\approx \left(c/\sqrt{\varepsilon^{(2)}_{yy}}\right)k_{1x}$. So,  new frequencies $\omega_2$ of waves that arise in WM~2 are not only uniformly proportional to spatial frequencies $k_{1x}$ but, for large values of $\psi$ correspond to new frequencies $\omega_2\sim k_{1x}$. Imagine that a complex source (also called \emph{object}) with strongly subwavelength details produces electromagnetic fields in front of the WM~1 sample. These subwavelength details will produce subwavelength oscillations in waves  propagating along the wires of WM~1, and high spatial frequencies $k_{1x}>k_{10}$ will carry out these oscillations so that the spatial spectrum of the propagating pulse repeats the spatial spectrum of the object field. 

Here, we should recall the story of far-field subwavelength imaging with WM which started from work~\cite{belov2005canalization}. Conceptual microwave studies (e.g.,~Refs.~\cite{belov2008transmission, zhao2010magnification}) stimulated the development of wire medium endoscopes {offering also imaging with strongly subwavelength resolution~\cite{tuniz2013metamaterial, brownless2015guided, casse2010super, zhou2010plasmon} besides the usual THz sensing}. In these endoscopes, subwavelength details of electric field distributions can be replicated at electrically long distances using plastic tubes filled with WM. The ultimate resolution of a WM endoscope is close to the period $a$ of the WM. For existing THz endoscopes $a\sim (0.05-0.1) \lambda$ \cite{tuniz2013metamaterial} but, in principle, this period can be much smaller \cite{silveirinha2008ultimate}. 
Both THz and infrared WM endoscopes suffer from restrictions related to the detection of subwavelength images. To capture the image, one needs THz detectors with the size of the order of $10\, \mu$m, whereas available far-field THz detectors are much larger \cite{tuniz2013metamaterial}. Near-ﬁeld THz detecting systems, which utilize microscopic probes offering subwavelength resolution, are based on raster scanning which makes them not suitable for in-situ measurements. The same and even tighter restrictions hold for infrared WM endoscopes. In the infrared experiments \cite{casse2010super, aleshire2020far, zhou2010plasmon}, subwavelength images were either recorded in a photoresist layer or measured indirectly with the use of photoluminescent markers. 

For THz endoscopes dedicated to in-situ imaging, one overcame the aforementioned limitations by developing a technology of 3D tapering~\cite{tuniz2013metamaterial}. With these conical WM endoscopes, it is possible to obtain magnified field distributions that couple to propagating waves, allowing the use of conventional THz optics for image acquisition. Such an endoscope is an optically thick multiwire ``cable'' of uniformly tapered wires. To practically achieve this uniform tapering, the wires are made of a liquid metal -- indium. The replacement of copper by indium results in rather high electric losses that restrict the length of the endoscope by approximately 1~cm~\cite{tuniz2013metamaterial}. For many THz applications, this length is insufficient. Moreover, it is comparable with the diameter of the conical base, which makes such endoscopes not enough flexible.      

The scenario of the temporal discontinuity that we have described here is an alternative approach to subwavelength imaging enabled by a WM. In this new approach, the spatially dispersive WM disappears being substituted by a split WM which is simply a uniaxial dielectric. The result of this temporal discontinuity is an angular pattern of propagating waves, where the frequencies and intensities  ``encode'' the spatial structure of the object. {As it is mentioned in Introduction}, the object image can be restored by a spectroscopic measurement of the angular pattern produced by the WM sample after the described temporal jump. 

To restore an arbitrary wave package propagating in WM~1 at a frequency $\omega_1$ it is enough to retrieve the amplitudes $E_1$ corresponding to the angles $\psi$ under which the time-refracted and time reflected waves propagate. First, from \eqref{final} we see that $E_1=E_2^{+}-E_2^{-}$. It means that we can find $|E_1|$ from measured amplitudes and phases of emitted time-refracted and time-reflected waves. However, measuring the phases of time-refracted and time-reflected waves is not necessary. {Really, for a given angle $\psi$ after measuring at the corresponding frequency $\omega_2$ we may find $|E_1|$ via the intensity either of the time-refracted ($I_{2+}$) wave or time-reflected ($I_{2-}$) wave. It is so because the relations \eqref{final} are real-valued.}  Using \eqref{final} one can find both the amplitude and phase of $E_1$, if we know the permittivity $\varepsilon^{(2)}_{yy}$. This real-valued (in the lossless approximation) permittivity can be found, e.g. detecting the polarization axis of the electric field $E_2^+$ or $E_2^-$ and using the relation ${\varepsilon^{(2)}_{yy}}=1/\tan\phi\tan\psi$. Thus, we will be able to restore the complex spatial spectrum of the spatial pulse that  propagated in WM~1 before the transition via standard spectroscopic/polarization measurements in the plane $(x-y)$.


\section{Outlook and Conclusions}

The revealed field transformation effect offers a possibility for 1D subwavelength imaging by measuring  TM-polarized plane waves that arise from  TEM-waves or quasi-TEM waves (in case of inductively loaded wires). The subwavelength details of the field distribution can be restored only along the axis $x$. If we rearrange the vectors $\_E_1$ and $\_H_1$ in Fig.~\ref{fig1} so that $\_E_1=\_z_0E_1$, and $\_H_1=\_x_0H_1$, the waves in WM~2 whose response is described by Eq.~\eqref{eps2} should be two ordinary waves. Their dispersion equation is simply $k_{2x}^2+k_{2y}^2=k_{20}^2$. Time boundary conditions for the wavevector \cite{morgenthaler1958velocity} in this case would give the solution $\omega_2=\omega_1\sqrt{1+\cot^2{\psi}}$. However, this solution is not compatible with the time boundary condition $\_B_1=\_B_2$. Since the permeability does not change, it means $H_1=\_x_0\cdot (\_H_{2}^++\_H_{2}^-)$. This condition cannot be satisfied with a pair of ordinary waves i.e. the components of the wave vector $k_{1y}$ and $k_{1x}$ cannot be preserved together with the vector $\_B$. It means that the model of the time interface from  \cite{morgenthaler1958velocity} does not hold. The same situation occurs if the wires of WM~1 disappear at $t=0$ and WM~2 transits to free space. 

The reason of the failure of the Morgenthaler approach \cite{morgenthaler1958velocity} for this case of the initial wave polarization is the  spatial dispersion in WM~1. In general, this approach is not valid for spatially dispersive media. The reason why for the pulse with the $x$-component of the spatial spectrum this approach fortunately turned out to be applicable is that an only pair of the time-refracted and time-reflected waves arises with an only new frequency $\omega_2$. In the dual case of the polarization, more than one new frequency arises, and we plan to study this case solving all equations in the time domain. 

At this stage, we did not consider the impact of the finite length of the wires and finite width of the real endoscope analyzing only a temporal discontinuity in an infinitely extended WM. The impact of the finite sizes cannot change the underlying physics of the found effect. Indeed, in future studies, these factors will be taken into account. The main goal of this paper was to initiate studies of time interfaces in spatially dispersive media and show that a temporal discontinuity in a WM opens the gate to a new spectroscopic method of 1D subwavelength imaging which can be implemented in practice using split wires initially connected by micro-discharges. 


\section*{Acknowledgement} 

This work was supported by the European Innovation Council program Pathfinder Open 2022 through the project PULSE, project number 101099313.


\bibliography{References} 

\begin{thebibliography}{52}%
\makeatletter
\providecommand \@ifxundefined [1]{%
 \@ifx{#1\undefined}
}%
\providecommand \@ifnum [1]{%
 \ifnum #1\expandafter \@firstoftwo
 \else \expandafter \@secondoftwo
 \fi
}%
\providecommand \@ifx [1]{%
 \ifx #1\expandafter \@firstoftwo
 \else \expandafter \@secondoftwo
 \fi
}%
\providecommand \natexlab [1]{#1}%
\providecommand \enquote  [1]{``#1''}%
\providecommand \bibnamefont  [1]{#1}%
\providecommand \bibfnamefont [1]{#1}%
\providecommand \citenamefont [1]{#1}%
\providecommand \href@noop [0]{\@secondoftwo}%
\providecommand \href [0]{\begingroup \@sanitize@url \@href}%
\providecommand \@href[1]{\@@startlink{#1}\@@href}%
\providecommand \@@href[1]{\endgroup#1\@@endlink}%
\providecommand \@sanitize@url [0]{\catcode `\\12\catcode `\$12\catcode `\&12\catcode `\#12\catcode `\^12\catcode `\_12\catcode `\%12\relax}%
\providecommand \@@startlink[1]{}%
\providecommand \@@endlink[0]{}%
\providecommand \url  [0]{\begingroup\@sanitize@url \@url }%
\providecommand \@url [1]{\endgroup\@href {#1}{\urlprefix }}%
\providecommand \urlprefix  [0]{URL }%
\providecommand \Eprint [0]{\href }%
\providecommand \doibase [0]{https://doi.org/}%
\providecommand \selectlanguage [0]{\@gobble}%
\providecommand \bibinfo  [0]{\@secondoftwo}%
\providecommand \bibfield  [0]{\@secondoftwo}%
\providecommand \translation [1]{[#1]}%
\providecommand \BibitemOpen [0]{}%
\providecommand \bibitemStop [0]{}%
\providecommand \bibitemNoStop [0]{.\EOS\space}%
\providecommand \EOS [0]{\spacefactor3000\relax}%
\providecommand \BibitemShut  [1]{\csname bibitem#1\endcsname}%
\let\auto@bib@innerbib\@empty
\bibitem [{\citenamefont {Galiffi}\ \emph {et~al.}(2022)\citenamefont {Galiffi}, \citenamefont {Tirole}, \citenamefont {Yin}, \citenamefont {Li}, \citenamefont {Vezzoli}, \citenamefont {Huidobro}, \citenamefont {Silveirinha}, \citenamefont {Sapienza}, \citenamefont {Al{\`u}},\ and\ \citenamefont {Pendry}}]{galiffi2022photonics}%
  \BibitemOpen
  \bibfield  {author} {\bibinfo {author} {\bibfnamefont {E.}~\bibnamefont {Galiffi}}, \bibinfo {author} {\bibfnamefont {R.}~\bibnamefont {Tirole}}, \bibinfo {author} {\bibfnamefont {S.}~\bibnamefont {Yin}}, \bibinfo {author} {\bibfnamefont {H.}~\bibnamefont {Li}}, \bibinfo {author} {\bibfnamefont {S.}~\bibnamefont {Vezzoli}}, \bibinfo {author} {\bibfnamefont {P.~A.}\ \bibnamefont {Huidobro}}, \bibinfo {author} {\bibfnamefont {M.~G.}\ \bibnamefont {Silveirinha}}, \bibinfo {author} {\bibfnamefont {R.}~\bibnamefont {Sapienza}}, \bibinfo {author} {\bibfnamefont {A.}~\bibnamefont {Al{\`u}}},\ and\ \bibinfo {author} {\bibfnamefont {J.}~\bibnamefont {Pendry}},\ }\bibfield  {title} {\bibinfo {title} {Photonics of time-varying media},\ }\href@noop {} {\bibfield  {journal} {\bibinfo  {journal} {Advanced Photonics}\ }\textbf {\bibinfo {volume} {4}},\ \bibinfo {pages} {014002} (\bibinfo {year} {2022})}\BibitemShut {NoStop}%
\bibitem [{\citenamefont {Ptitcyn}\ \emph {et~al.}(2023)\citenamefont {Ptitcyn}, \citenamefont {Mirmoosa}, \citenamefont {Sotoodehfar},\ and\ \citenamefont {Tretyakov}}]{ptitcyn2023tutorial}%
  \BibitemOpen
  \bibfield  {author} {\bibinfo {author} {\bibfnamefont {G.}~\bibnamefont {Ptitcyn}}, \bibinfo {author} {\bibfnamefont {M.~S.}\ \bibnamefont {Mirmoosa}}, \bibinfo {author} {\bibfnamefont {A.}~\bibnamefont {Sotoodehfar}},\ and\ \bibinfo {author} {\bibfnamefont {S.~A.}\ \bibnamefont {Tretyakov}},\ }\bibfield  {title} {\bibinfo {title} {A tutorial on the basics of time-varying electromagnetic systems and circuits: Historic overview and basic concepts of time-modulation},\ }\href@noop {} {\bibfield  {journal} {\bibinfo  {journal} {IEEE Antennas and Propagation Magazine}\ }\textbf {\bibinfo {volume} {65}},\ \bibinfo {pages} {10} (\bibinfo {year} {2023})}\BibitemShut {NoStop}%
\bibitem [{\citenamefont {Mostafa}\ \emph {et~al.}(2024)\citenamefont {Mostafa}, \citenamefont {Mirmoosa}, \citenamefont {Sidorenko}, \citenamefont {Asadchy},\ and\ \citenamefont {Tretyakov}}]{mostafa2024temporal}%
  \BibitemOpen
  \bibfield  {author} {\bibinfo {author} {\bibfnamefont {M.}~\bibnamefont {Mostafa}}, \bibinfo {author} {\bibfnamefont {M.}~\bibnamefont {Mirmoosa}}, \bibinfo {author} {\bibfnamefont {M.}~\bibnamefont {Sidorenko}}, \bibinfo {author} {\bibfnamefont {V.}~\bibnamefont {Asadchy}},\ and\ \bibinfo {author} {\bibfnamefont {S.}~\bibnamefont {Tretyakov}},\ }\bibfield  {title} {\bibinfo {title} {Temporal interfaces in complex electromagnetic materials: an overview},\ }\href@noop {} {\bibfield  {journal} {\bibinfo  {journal} {Optical Materials Express}\ }\textbf {\bibinfo {volume} {14}},\ \bibinfo {pages} {1103} (\bibinfo {year} {2024})}\BibitemShut {NoStop}%
\bibitem [{\citenamefont {Morgenthaler}(1958)}]{morgenthaler1958velocity}%
  \BibitemOpen
  \bibfield  {author} {\bibinfo {author} {\bibfnamefont {F.~R.}\ \bibnamefont {Morgenthaler}},\ }\bibfield  {title} {\bibinfo {title} {Velocity modulation of electromagnetic waves},\ }\href@noop {} {\bibfield  {journal} {\bibinfo  {journal} {IRE Transactions on Microwave Theory and Techniques}\ }\textbf {\bibinfo {volume} {6}},\ \bibinfo {pages} {167} (\bibinfo {year} {1958})}\BibitemShut {NoStop}%
\bibitem [{\citenamefont {Mendon{\c{c}}a}\ and\ \citenamefont {Shukla}(2002)}]{mendoncca2002time}%
  \BibitemOpen
  \bibfield  {author} {\bibinfo {author} {\bibfnamefont {J.}~\bibnamefont {Mendon{\c{c}}a}}\ and\ \bibinfo {author} {\bibfnamefont {P.}~\bibnamefont {Shukla}},\ }\bibfield  {title} {\bibinfo {title} {Time refraction and time reflection: two basic concepts},\ }\href@noop {} {\bibfield  {journal} {\bibinfo  {journal} {Physica Scripta}\ }\textbf {\bibinfo {volume} {65}},\ \bibinfo {pages} {160} (\bibinfo {year} {2002})}\BibitemShut {NoStop}%
\bibitem [{\citenamefont {Xiao}\ \emph {et~al.}(2014)\citenamefont {Xiao}, \citenamefont {Maywar},\ and\ \citenamefont {Agrawal}}]{Agrawal2014RTC}%
  \BibitemOpen
  \bibfield  {author} {\bibinfo {author} {\bibfnamefont {Y.}~\bibnamefont {Xiao}}, \bibinfo {author} {\bibfnamefont {D.~N.}\ \bibnamefont {Maywar}},\ and\ \bibinfo {author} {\bibfnamefont {G.~P.}\ \bibnamefont {Agrawal}},\ }\bibfield  {title} {\bibinfo {title} {Reflection and transmission of electromagnetic waves at a temporal boundary},\ }\href@noop {} {\bibfield  {journal} {\bibinfo  {journal} {Optics letters}\ }\textbf {\bibinfo {volume} {39}},\ \bibinfo {pages} {574} (\bibinfo {year} {2014})}\BibitemShut {NoStop}%
\bibitem [{\citenamefont {Mendon{\c{c}}a}\ \emph {et~al.}(2000)\citenamefont {Mendon{\c{c}}a}, \citenamefont {Guerreiro},\ and\ \citenamefont {Martins}}]{mendoncca2000quantum}%
  \BibitemOpen
  \bibfield  {author} {\bibinfo {author} {\bibfnamefont {J.}~\bibnamefont {Mendon{\c{c}}a}}, \bibinfo {author} {\bibfnamefont {A.}~\bibnamefont {Guerreiro}},\ and\ \bibinfo {author} {\bibfnamefont {A.~M.}\ \bibnamefont {Martins}},\ }\bibfield  {title} {\bibinfo {title} {Quantum theory of time refraction},\ }\href@noop {} {\bibfield  {journal} {\bibinfo  {journal} {Physical Review A}\ }\textbf {\bibinfo {volume} {62}},\ \bibinfo {pages} {033805} (\bibinfo {year} {2000})}\BibitemShut {NoStop}%
\bibitem [{\citenamefont {Mirmoosa}\ \emph {et~al.}(2023)\citenamefont {Mirmoosa}, \citenamefont {Set{\"a}l{\"a}},\ and\ \citenamefont {Norrman}}]{mirmoosa2023quantum}%
  \BibitemOpen
  \bibfield  {author} {\bibinfo {author} {\bibfnamefont {M.}~\bibnamefont {Mirmoosa}}, \bibinfo {author} {\bibfnamefont {T.}~\bibnamefont {Set{\"a}l{\"a}}},\ and\ \bibinfo {author} {\bibfnamefont {A.}~\bibnamefont {Norrman}},\ }\bibfield  {title} {\bibinfo {title} {Quantum theory of wave scattering from electromagnetic time interfaces},\ }\href@noop {} {\bibfield  {journal} {\bibinfo  {journal} {arXiv preprint arXiv:2312.15178}\ } (\bibinfo {year} {2023})}\BibitemShut {NoStop}%
\bibitem [{\citenamefont {Wilks}\ \emph {et~al.}(1988)\citenamefont {Wilks}, \citenamefont {Dawson},\ and\ \citenamefont {Mori}}]{Wilks1988TH}%
  \BibitemOpen
  \bibfield  {author} {\bibinfo {author} {\bibfnamefont {S.~C.}\ \bibnamefont {Wilks}}, \bibinfo {author} {\bibfnamefont {J.~M.}\ \bibnamefont {Dawson}},\ and\ \bibinfo {author} {\bibfnamefont {W.~B.}\ \bibnamefont {Mori}},\ }\bibfield  {title} {\bibinfo {title} {Frequency up-conversion of electromagnetic radiation with use of an overdense plasma},\ }\href {https://doi.org/10.1103/PhysRevLett.61.337} {\bibfield  {journal} {\bibinfo  {journal} {Phys. Rev. Lett.}\ }\textbf {\bibinfo {volume} {61}},\ \bibinfo {pages} {337} (\bibinfo {year} {1988})}\BibitemShut {NoStop}%
\bibitem [{\citenamefont {Yugami}\ \emph {et~al.}(2002)\citenamefont {Yugami}, \citenamefont {Niiyama}, \citenamefont {Higashiguchi}, \citenamefont {Gao}, \citenamefont {Sasaki}, \citenamefont {Ito},\ and\ \citenamefont {Nishida}}]{Yugami2002EXPER}%
  \BibitemOpen
  \bibfield  {author} {\bibinfo {author} {\bibfnamefont {N.}~\bibnamefont {Yugami}}, \bibinfo {author} {\bibfnamefont {T.}~\bibnamefont {Niiyama}}, \bibinfo {author} {\bibfnamefont {T.}~\bibnamefont {Higashiguchi}}, \bibinfo {author} {\bibfnamefont {H.}~\bibnamefont {Gao}}, \bibinfo {author} {\bibfnamefont {S.}~\bibnamefont {Sasaki}}, \bibinfo {author} {\bibfnamefont {H.}~\bibnamefont {Ito}},\ and\ \bibinfo {author} {\bibfnamefont {Y.}~\bibnamefont {Nishida}},\ }\bibfield  {title} {\bibinfo {title} {Experimental observation of short-pulse upshifted frequency microwaves from a laser-created overdense plasma},\ }\href@noop {} {\bibfield  {journal} {\bibinfo  {journal} {Phys. Rev. E}\ }\textbf {\bibinfo {volume} {65}},\ \bibinfo {pages} {036505} (\bibinfo {year} {2002})}\BibitemShut {NoStop}%
\bibitem [{\citenamefont {Nishida}\ \emph {et~al.}(2012)\citenamefont {Nishida}, \citenamefont {Yugami}, \citenamefont {Higashiguchi}, \citenamefont {Otsuka}, \citenamefont {Suzuki}, \citenamefont {Nakata}, \citenamefont {Sentoku},\ and\ \citenamefont {Kodama}}]{nishida2012EXP}%
  \BibitemOpen
  \bibfield  {author} {\bibinfo {author} {\bibfnamefont {A.}~\bibnamefont {Nishida}}, \bibinfo {author} {\bibfnamefont {N.}~\bibnamefont {Yugami}}, \bibinfo {author} {\bibfnamefont {T.}~\bibnamefont {Higashiguchi}}, \bibinfo {author} {\bibfnamefont {T.}~\bibnamefont {Otsuka}}, \bibinfo {author} {\bibfnamefont {F.}~\bibnamefont {Suzuki}}, \bibinfo {author} {\bibfnamefont {M.}~\bibnamefont {Nakata}}, \bibinfo {author} {\bibfnamefont {Y.}~\bibnamefont {Sentoku}},\ and\ \bibinfo {author} {\bibfnamefont {R.}~\bibnamefont {Kodama}},\ }\bibfield  {title} {\bibinfo {title} {Experimental observation of frequency up-conversion by flash ionization},\ }\href@noop {} {\bibfield  {journal} {\bibinfo  {journal} {Applied Physics Letters}\ }\textbf {\bibinfo {volume} {101}} (\bibinfo {year} {2012})}\BibitemShut {NoStop}%
\bibitem [{\citenamefont {Bacot}\ \emph {et~al.}(2016)\citenamefont {Bacot}, \citenamefont {Labousse}, \citenamefont {Eddi}, \citenamefont {Fink},\ and\ \citenamefont {Fort}}]{water}%
  \BibitemOpen
  \bibfield  {author} {\bibinfo {author} {\bibfnamefont {V.}~\bibnamefont {Bacot}}, \bibinfo {author} {\bibfnamefont {M.}~\bibnamefont {Labousse}}, \bibinfo {author} {\bibfnamefont {A.}~\bibnamefont {Eddi}}, \bibinfo {author} {\bibfnamefont {M.}~\bibnamefont {Fink}},\ and\ \bibinfo {author} {\bibfnamefont {E.}~\bibnamefont {Fort}},\ }\bibfield  {title} {\bibinfo {title} {Time reversal and holography with spacetime transformations},\ }\href@noop {} {\bibfield  {journal} {\bibinfo  {journal} {Nature Physics}\ }\textbf {\bibinfo {volume} {12}},\ \bibinfo {pages} {972} (\bibinfo {year} {2016})}\BibitemShut {NoStop}%
\bibitem [{\citenamefont {Moussa}\ \emph {et~al.}(2023)\citenamefont {Moussa}, \citenamefont {Xu}, \citenamefont {Yin}, \citenamefont {Galiffi}, \citenamefont {Ra’di},\ and\ \citenamefont {Alù}}]{Alu_TL}%
  \BibitemOpen
  \bibfield  {author} {\bibinfo {author} {\bibfnamefont {H.}~\bibnamefont {Moussa}}, \bibinfo {author} {\bibfnamefont {G.}~\bibnamefont {Xu}}, \bibinfo {author} {\bibfnamefont {S.}~\bibnamefont {Yin}}, \bibinfo {author} {\bibfnamefont {E.}~\bibnamefont {Galiffi}}, \bibinfo {author} {\bibfnamefont {Y.}~\bibnamefont {Ra’di}},\ and\ \bibinfo {author} {\bibfnamefont {A.}~\bibnamefont {Alù}},\ }\bibfield  {title} {\bibinfo {title} {Observation of temporal reflection and broadband frequency translation at photonic time interfaces},\ }\href {https://doi.org/10.1038/s41567-023-01975-y} {\bibfield  {journal} {\bibinfo  {journal} {Nature Physics}\ }\textbf {\bibinfo {volume} {19}},\ \bibinfo {pages} {863} (\bibinfo {year} {2023})}\BibitemShut {NoStop}%
\bibitem [{\citenamefont {Xu}\ \emph {et~al.}(2021)\citenamefont {Xu}, \citenamefont {Mai},\ and\ \citenamefont {Werner}}]{xu2021complete}%
  \BibitemOpen
  \bibfield  {author} {\bibinfo {author} {\bibfnamefont {J.}~\bibnamefont {Xu}}, \bibinfo {author} {\bibfnamefont {W.}~\bibnamefont {Mai}},\ and\ \bibinfo {author} {\bibfnamefont {D.~H.}\ \bibnamefont {Werner}},\ }\bibfield  {title} {\bibinfo {title} {Complete polarization conversion using anisotropic temporal slabs},\ }\href@noop {} {\bibfield  {journal} {\bibinfo  {journal} {Optics Letters}\ }\textbf {\bibinfo {volume} {46}},\ \bibinfo {pages} {1373} (\bibinfo {year} {2021})}\BibitemShut {NoStop}%
\bibitem [{\citenamefont {Rizza}\ \emph {et~al.}(2023)\citenamefont {Rizza}, \citenamefont {Castaldi},\ and\ \citenamefont {Galdi}}]{RizzaCastaldiGaldi23}%
  \BibitemOpen
  \bibfield  {author} {\bibinfo {author} {\bibfnamefont {C.}~\bibnamefont {Rizza}}, \bibinfo {author} {\bibfnamefont {G.}~\bibnamefont {Castaldi}},\ and\ \bibinfo {author} {\bibfnamefont {V.}~\bibnamefont {Galdi}},\ }\bibfield  {title} {\bibinfo {title} {Spin-controlled photonics via temporal anisotropy},\ }\href {https://doi.org/doi:10.1515/nanoph-2022-0809} {\bibfield  {journal} {\bibinfo  {journal} {Nanophotonics}\ }\textbf {\bibinfo {volume} {12}},\ \bibinfo {pages} {2891} (\bibinfo {year} {2023})}\BibitemShut {NoStop}%
\bibitem [{\citenamefont {Mostafa}\ \emph {et~al.}(2023)\citenamefont {Mostafa}, \citenamefont {Mirmoosa},\ and\ \citenamefont {Tretyakov}}]{mostafa2023spin}%
  \BibitemOpen
  \bibfield  {author} {\bibinfo {author} {\bibfnamefont {M.~H.~M.}\ \bibnamefont {Mostafa}}, \bibinfo {author} {\bibfnamefont {M.~S.}\ \bibnamefont {Mirmoosa}},\ and\ \bibinfo {author} {\bibfnamefont {S.~A.}\ \bibnamefont {Tretyakov}},\ }\bibfield  {title} {\bibinfo {title} {Spin-dependent phenomena at chiral temporal interfaces},\ }\href {https://doi.org/doi:10.1515/nanoph-2022-0805} {\bibfield  {journal} {\bibinfo  {journal} {Nanophotonics}\ }\textbf {\bibinfo {volume} {12}},\ \bibinfo {pages} {2881} (\bibinfo {year} {2023})}\BibitemShut {NoStop}%
\bibitem [{\citenamefont {Mirmoosa}\ \emph {et~al.}(2024)\citenamefont {Mirmoosa}, \citenamefont {Mostafa}, \citenamefont {Norrman},\ and\ \citenamefont {Tretyakov}}]{mirmoosa2024time}%
  \BibitemOpen
  \bibfield  {author} {\bibinfo {author} {\bibfnamefont {M.}~\bibnamefont {Mirmoosa}}, \bibinfo {author} {\bibfnamefont {M.}~\bibnamefont {Mostafa}}, \bibinfo {author} {\bibfnamefont {A.}~\bibnamefont {Norrman}},\ and\ \bibinfo {author} {\bibfnamefont {S.}~\bibnamefont {Tretyakov}},\ }\bibfield  {title} {\bibinfo {title} {Time interfaces in bianisotropic media},\ }\href@noop {} {\bibfield  {journal} {\bibinfo  {journal} {Physical Review Research}\ }\textbf {\bibinfo {volume} {6}},\ \bibinfo {pages} {013334} (\bibinfo {year} {2024})}\BibitemShut {NoStop}%
\bibitem [{\citenamefont {Akbarzadeh}\ \emph {et~al.}(2018)\citenamefont {Akbarzadeh}, \citenamefont {Chamanara},\ and\ \citenamefont {Caloz}}]{akbarzadeh2018inverse}%
  \BibitemOpen
  \bibfield  {author} {\bibinfo {author} {\bibfnamefont {A.}~\bibnamefont {Akbarzadeh}}, \bibinfo {author} {\bibfnamefont {N.}~\bibnamefont {Chamanara}},\ and\ \bibinfo {author} {\bibfnamefont {C.}~\bibnamefont {Caloz}},\ }\bibfield  {title} {\bibinfo {title} {Inverse prism based on temporal discontinuity and spatial dispersion},\ }\href@noop {} {\bibfield  {journal} {\bibinfo  {journal} {Optics Letters}\ }\textbf {\bibinfo {volume} {43}},\ \bibinfo {pages} {3297} (\bibinfo {year} {2018})}\BibitemShut {NoStop}%
\bibitem [{\citenamefont {Ramaccia}\ \emph {et~al.}(2020)\citenamefont {Ramaccia}, \citenamefont {Toscano},\ and\ \citenamefont {Bilotti}}]{ramaccia2020light}%
  \BibitemOpen
  \bibfield  {author} {\bibinfo {author} {\bibfnamefont {D.}~\bibnamefont {Ramaccia}}, \bibinfo {author} {\bibfnamefont {A.}~\bibnamefont {Toscano}},\ and\ \bibinfo {author} {\bibfnamefont {F.}~\bibnamefont {Bilotti}},\ }\bibfield  {title} {\bibinfo {title} {Light propagation through metamaterial temporal slabs: reflection, refraction, and special cases},\ }\href@noop {} {\bibfield  {journal} {\bibinfo  {journal} {Optics Letters}\ }\textbf {\bibinfo {volume} {45}},\ \bibinfo {pages} {5836} (\bibinfo {year} {2020})}\BibitemShut {NoStop}%
\bibitem [{\citenamefont {Pacheco-Pe{\~n}a}\ and\ \citenamefont {Engheta}(2020{\natexlab{a}})}]{pacheco2020anti}%
  \BibitemOpen
  \bibfield  {author} {\bibinfo {author} {\bibfnamefont {V.}~\bibnamefont {Pacheco-Pe{\~n}a}}\ and\ \bibinfo {author} {\bibfnamefont {N.}~\bibnamefont {Engheta}},\ }\bibfield  {title} {\bibinfo {title} {Antireflection temporal coatings},\ }\href@noop {} {\bibfield  {journal} {\bibinfo  {journal} {Optica}\ }\textbf {\bibinfo {volume} {7}},\ \bibinfo {pages} {323} (\bibinfo {year} {2020}{\natexlab{a}})}\BibitemShut {NoStop}%
\bibitem [{\citenamefont {Pacheco-Pe{\~n}a}\ and\ \citenamefont {Engheta}(2020{\natexlab{b}})}]{pacheco2020aiming}%
  \BibitemOpen
  \bibfield  {author} {\bibinfo {author} {\bibfnamefont {V.}~\bibnamefont {Pacheco-Pe{\~n}a}}\ and\ \bibinfo {author} {\bibfnamefont {N.}~\bibnamefont {Engheta}},\ }\bibfield  {title} {\bibinfo {title} {Temporal aiming},\ }\href@noop {} {\bibfield  {journal} {\bibinfo  {journal} {Light: Science and Applications}\ }\textbf {\bibinfo {volume} {9}},\ \bibinfo {pages} {129} (\bibinfo {year} {2020}{\natexlab{b}})}\BibitemShut {NoStop}%
\bibitem [{\citenamefont {Wang}\ \emph {et~al.}(2023)\citenamefont {Wang}, \citenamefont {Mirmoosa},\ and\ \citenamefont {Tretyakov}}]{wang2023controlling}%
  \BibitemOpen
  \bibfield  {author} {\bibinfo {author} {\bibfnamefont {X.}~\bibnamefont {Wang}}, \bibinfo {author} {\bibfnamefont {M.~S.}\ \bibnamefont {Mirmoosa}},\ and\ \bibinfo {author} {\bibfnamefont {S.~A.}\ \bibnamefont {Tretyakov}},\ }\bibfield  {title} {\bibinfo {title} {Controlling surface waves with temporal discontinuities of metasurfaces},\ }\href {https://doi.org/doi:10.1515/nanoph-2022-0685} {\bibfield  {journal} {\bibinfo  {journal} {Nanophotonics}\ }\textbf {\bibinfo {volume} {12}},\ \bibinfo {pages} {2813} (\bibinfo {year} {2023})}\BibitemShut {NoStop}%
\bibitem [{\citenamefont {Jiang}(1975)}]{jiang1975wave}%
  \BibitemOpen
  \bibfield  {author} {\bibinfo {author} {\bibfnamefont {C.-L.}\ \bibnamefont {Jiang}},\ }\bibfield  {title} {\bibinfo {title} {Wave propagation and dipole radiation in a suddenly created plasma},\ }\href@noop {} {\bibfield  {journal} {\bibinfo  {journal} {IEEE Transactions on Antennas and Propagation}\ }\textbf {\bibinfo {volume} {23}},\ \bibinfo {pages} {83} (\bibinfo {year} {1975})}\BibitemShut {NoStop}%
\bibitem [{\citenamefont {Kalluri}(1988)}]{kalluri1988reflection}%
  \BibitemOpen
  \bibfield  {author} {\bibinfo {author} {\bibfnamefont {D.~K.}\ \bibnamefont {Kalluri}},\ }\bibfield  {title} {\bibinfo {title} {On reflection from a suddenly created plasma half-space: Transient solution},\ }\href@noop {} {\bibfield  {journal} {\bibinfo  {journal} {IEEE transactions on plasma science}\ }\textbf {\bibinfo {volume} {16}},\ \bibinfo {pages} {11} (\bibinfo {year} {1988})}\BibitemShut {NoStop}%
\bibitem [{\citenamefont {Shirokova}\ \emph {et~al.}(2019)\citenamefont {Shirokova}, \citenamefont {Maslov},\ and\ \citenamefont {Bakunov}}]{Grap19SPP}%
  \BibitemOpen
  \bibfield  {author} {\bibinfo {author} {\bibfnamefont {A.~V.}\ \bibnamefont {Shirokova}}, \bibinfo {author} {\bibfnamefont {A.~V.}\ \bibnamefont {Maslov}},\ and\ \bibinfo {author} {\bibfnamefont {M.~I.}\ \bibnamefont {Bakunov}},\ }\bibfield  {title} {\bibinfo {title} {Scattering of surface plasmons on graphene by abrupt free-carrier generation},\ }\href {https://doi.org/10.1103/PhysRevB.100.045424} {\bibfield  {journal} {\bibinfo  {journal} {Phys. Rev. B}\ }\textbf {\bibinfo {volume} {100}},\ \bibinfo {pages} {045424} (\bibinfo {year} {2019})}\BibitemShut {NoStop}%
\bibitem [{\citenamefont {Simovski}\ \emph {et~al.}(2012)\citenamefont {Simovski}, \citenamefont {Belov}, \citenamefont {Atrashchenko},\ and\ \citenamefont {Kivshar}}]{simovski2012wire}%
  \BibitemOpen
  \bibfield  {author} {\bibinfo {author} {\bibfnamefont {C.~R.}\ \bibnamefont {Simovski}}, \bibinfo {author} {\bibfnamefont {P.~A.}\ \bibnamefont {Belov}}, \bibinfo {author} {\bibfnamefont {A.~V.}\ \bibnamefont {Atrashchenko}},\ and\ \bibinfo {author} {\bibfnamefont {Y.~S.}\ \bibnamefont {Kivshar}},\ }\bibfield  {title} {\bibinfo {title} {Wire metamaterials: physics and applications},\ }\href@noop {} {\bibfield  {journal} {\bibinfo  {journal} {Advanced materials}\ }\textbf {\bibinfo {volume} {24}},\ \bibinfo {pages} {4229} (\bibinfo {year} {2012})}\BibitemShut {NoStop}%
\bibitem [{\citenamefont {Slobozhanyuk}\ \emph {et~al.}(2014)\citenamefont {Slobozhanyuk}, \citenamefont {Kozachenko}, \citenamefont {Melchakova}, \citenamefont {Simovski},\ and\ \citenamefont {Belov}}]{Slobo}%
  \BibitemOpen
  \bibfield  {author} {\bibinfo {author} {\bibfnamefont {A.}~\bibnamefont {Slobozhanyuk}}, \bibinfo {author} {\bibfnamefont {A.}~\bibnamefont {Kozachenko}}, \bibinfo {author} {\bibfnamefont {I.}~\bibnamefont {Melchakova}}, \bibinfo {author} {\bibfnamefont {C.}~\bibnamefont {Simovski}},\ and\ \bibinfo {author} {\bibfnamefont {P.}~\bibnamefont {Belov}},\ }\bibfield  {title} {\bibinfo {title} {Endoscopes based on ultimately anisotropic metamaterials for magnetic resonance tomography},\ }\href@noop {} {\bibfield  {journal} {\bibinfo  {journal} {Journal of Communications Technology and Electronics}\ }\textbf {\bibinfo {volume} {59}},\ \bibinfo {pages} {562} (\bibinfo {year} {2014})}\BibitemShut {NoStop}%
\bibitem [{\citenamefont {Belov}\ \emph {et~al.}(2005)\citenamefont {Belov}, \citenamefont {Simovski},\ and\ \citenamefont {Ikonen}}]{belov2005canalization}%
  \BibitemOpen
  \bibfield  {author} {\bibinfo {author} {\bibfnamefont {P.~A.}\ \bibnamefont {Belov}}, \bibinfo {author} {\bibfnamefont {C.~R.}\ \bibnamefont {Simovski}},\ and\ \bibinfo {author} {\bibfnamefont {P.}~\bibnamefont {Ikonen}},\ }\bibfield  {title} {\bibinfo {title} {Canalization of subwavelength images by electromagnetic crystals},\ }\href@noop {} {\bibfield  {journal} {\bibinfo  {journal} {Physical Review B}\ }\textbf {\bibinfo {volume} {71}},\ \bibinfo {pages} {193105} (\bibinfo {year} {2005})}\BibitemShut {NoStop}%
\bibitem [{\citenamefont {Belov}\ \emph {et~al.}(2008)\citenamefont {Belov}, \citenamefont {Zhao}, \citenamefont {Tse}, \citenamefont {Ikonen}, \citenamefont {Silveirinha}, \citenamefont {Simovski}, \citenamefont {Tretyakov}, \citenamefont {Hao},\ and\ \citenamefont {Parini}}]{belov2008transmission}%
  \BibitemOpen
  \bibfield  {author} {\bibinfo {author} {\bibfnamefont {P.~A.}\ \bibnamefont {Belov}}, \bibinfo {author} {\bibfnamefont {Y.}~\bibnamefont {Zhao}}, \bibinfo {author} {\bibfnamefont {S.}~\bibnamefont {Tse}}, \bibinfo {author} {\bibfnamefont {P.}~\bibnamefont {Ikonen}}, \bibinfo {author} {\bibfnamefont {M.~G.}\ \bibnamefont {Silveirinha}}, \bibinfo {author} {\bibfnamefont {C.~R.}\ \bibnamefont {Simovski}}, \bibinfo {author} {\bibfnamefont {S.}~\bibnamefont {Tretyakov}}, \bibinfo {author} {\bibfnamefont {Y.}~\bibnamefont {Hao}},\ and\ \bibinfo {author} {\bibfnamefont {C.}~\bibnamefont {Parini}},\ }\bibfield  {title} {\bibinfo {title} {Transmission of images with subwavelength resolution to distances of several wavelengths in the microwave range},\ }\href@noop {} {\bibfield  {journal} {\bibinfo  {journal} {Physical Review B}\ }\textbf {\bibinfo {volume} {77}},\ \bibinfo {pages} {193108} (\bibinfo {year} {2008})}\BibitemShut {NoStop}%
\bibitem [{\citenamefont {Tuniz}\ \emph {et~al.}(2013)\citenamefont {Tuniz}, \citenamefont {Kaltenecker}, \citenamefont {Fischer}, \citenamefont {Walther}, \citenamefont {Fleming}, \citenamefont {Argyros},\ and\ \citenamefont {Kuhlmey}}]{tuniz2013metamaterial}%
  \BibitemOpen
  \bibfield  {author} {\bibinfo {author} {\bibfnamefont {A.}~\bibnamefont {Tuniz}}, \bibinfo {author} {\bibfnamefont {K.~J.}\ \bibnamefont {Kaltenecker}}, \bibinfo {author} {\bibfnamefont {B.~M.}\ \bibnamefont {Fischer}}, \bibinfo {author} {\bibfnamefont {M.}~\bibnamefont {Walther}}, \bibinfo {author} {\bibfnamefont {S.~C.}\ \bibnamefont {Fleming}}, \bibinfo {author} {\bibfnamefont {A.}~\bibnamefont {Argyros}},\ and\ \bibinfo {author} {\bibfnamefont {B.~T.}\ \bibnamefont {Kuhlmey}},\ }\bibfield  {title} {\bibinfo {title} {Metamaterial fibres for subdiffraction imaging and focusing at terahertz frequencies over optically long distances},\ }\href@noop {} {\bibfield  {journal} {\bibinfo  {journal} {Nature communications}\ }\textbf {\bibinfo {volume} {4}},\ \bibinfo {pages} {1} (\bibinfo {year} {2013})}\BibitemShut {NoStop}%
\bibitem [{\citenamefont {Casse}\ \emph {et~al.}(2010)\citenamefont {Casse}, \citenamefont {Lu}, \citenamefont {Huang}, \citenamefont {Gultepe}, \citenamefont {Menon},\ and\ \citenamefont {Sridhar}}]{casse2010super}%
  \BibitemOpen
  \bibfield  {author} {\bibinfo {author} {\bibfnamefont {B.}~\bibnamefont {Casse}}, \bibinfo {author} {\bibfnamefont {W.}~\bibnamefont {Lu}}, \bibinfo {author} {\bibfnamefont {Y.}~\bibnamefont {Huang}}, \bibinfo {author} {\bibfnamefont {E.}~\bibnamefont {Gultepe}}, \bibinfo {author} {\bibfnamefont {L.}~\bibnamefont {Menon}},\ and\ \bibinfo {author} {\bibfnamefont {S.}~\bibnamefont {Sridhar}},\ }\bibfield  {title} {\bibinfo {title} {Super-resolution imaging using a three-dimensional metamaterials nanolens},\ }\href@noop {} {\bibfield  {journal} {\bibinfo  {journal} {Applied Physics Letters}\ }\textbf {\bibinfo {volume} {96}} (\bibinfo {year} {2010})}\BibitemShut {NoStop}%
\bibitem [{\citenamefont {Simovski}\ \emph {et~al.}(2015)\citenamefont {Simovski}, \citenamefont {Maslovski}, \citenamefont {Nefedov}, \citenamefont {Kosulnikov}, \citenamefont {Belov},\ and\ \citenamefont {Tretyakov}}]{hypersim}%
  \BibitemOpen
  \bibfield  {author} {\bibinfo {author} {\bibfnamefont {C.}~\bibnamefont {Simovski}}, \bibinfo {author} {\bibfnamefont {S.}~\bibnamefont {Maslovski}}, \bibinfo {author} {\bibfnamefont {I.}~\bibnamefont {Nefedov}}, \bibinfo {author} {\bibfnamefont {S.}~\bibnamefont {Kosulnikov}}, \bibinfo {author} {\bibfnamefont {P.}~\bibnamefont {Belov}},\ and\ \bibinfo {author} {\bibfnamefont {S.}~\bibnamefont {Tretyakov}},\ }\bibfield  {title} {\bibinfo {title} {Hyperlens makes thermal emission strongly super-planckian},\ }\href {https://doi.org/https://doi.org/10.1016/j.photonics.2014.12.005} {\bibfield  {journal} {\bibinfo  {journal} {Photonics and Nanostructures - Fundamentals and Applications}\ }\textbf {\bibinfo {volume} {13}},\ \bibinfo {pages} {31} (\bibinfo {year} {2015})}\BibitemShut {NoStop}%
\bibitem [{\citenamefont {Kosulnikov}\ \emph {et~al.}(2015)\citenamefont {Kosulnikov}, \citenamefont {Filonov}, \citenamefont {Glybovski}, \citenamefont {Belov}, \citenamefont {Tretyakov},\ and\ \citenamefont {Simovski}}]{Kos1}%
  \BibitemOpen
  \bibfield  {author} {\bibinfo {author} {\bibfnamefont {S.}~\bibnamefont {Kosulnikov}}, \bibinfo {author} {\bibfnamefont {D.}~\bibnamefont {Filonov}}, \bibinfo {author} {\bibfnamefont {S.}~\bibnamefont {Glybovski}}, \bibinfo {author} {\bibfnamefont {P.}~\bibnamefont {Belov}}, \bibinfo {author} {\bibfnamefont {S.}~\bibnamefont {Tretyakov}},\ and\ \bibinfo {author} {\bibfnamefont {C.}~\bibnamefont {Simovski}},\ }\bibfield  {title} {\bibinfo {title} {Wire-medium hyperlens for enhancing radiation from subwavelength dipole sources},\ }\href@noop {} {\bibfield  {journal} {\bibinfo  {journal} {IEEE Transactions on Antennas and Propagation}\ }\textbf {\bibinfo {volume} {63}},\ \bibinfo {pages} {4848} (\bibinfo {year} {2015})}\BibitemShut {NoStop}%
\bibitem [{\citenamefont {Kosulnikov}\ \emph {et~al.}(2016)\citenamefont {Kosulnikov}, \citenamefont {Vovchuk}, \citenamefont {Mirmoosa}, \citenamefont {Tretyakov}, \citenamefont {Glybovski},\ and\ \citenamefont {Simovski}}]{Kos2}%
  \BibitemOpen
  \bibfield  {author} {\bibinfo {author} {\bibfnamefont {S.~Y.}\ \bibnamefont {Kosulnikov}}, \bibinfo {author} {\bibfnamefont {D.}~\bibnamefont {Vovchuk}}, \bibinfo {author} {\bibfnamefont {M.}~\bibnamefont {Mirmoosa}}, \bibinfo {author} {\bibfnamefont {S.}~\bibnamefont {Tretyakov}}, \bibinfo {author} {\bibfnamefont {S.}~\bibnamefont {Glybovski}},\ and\ \bibinfo {author} {\bibfnamefont {C.}~\bibnamefont {Simovski}},\ }\bibfield  {title} {\bibinfo {title} {Enhancement of radiation with irregular wire media},\ }\href@noop {} {\bibfield  {journal} {\bibinfo  {journal} {IEEE Transactions on Antennas and Propagation}\ }\textbf {\bibinfo {volume} {64}},\ \bibinfo {pages} {5469} (\bibinfo {year} {2016})}\BibitemShut {NoStop}%
\bibitem [{\citenamefont {Belov}\ \emph {et~al.}(2003)\citenamefont {Belov}, \citenamefont {Marques}, \citenamefont {Maslovski}, \citenamefont {Nefedov}, \citenamefont {Silveirinha}, \citenamefont {Simovski},\ and\ \citenamefont {Tretyakov}}]{belov2003strong}%
  \BibitemOpen
  \bibfield  {author} {\bibinfo {author} {\bibfnamefont {P.~A.}\ \bibnamefont {Belov}}, \bibinfo {author} {\bibfnamefont {R.}~\bibnamefont {Marques}}, \bibinfo {author} {\bibfnamefont {S.~I.}\ \bibnamefont {Maslovski}}, \bibinfo {author} {\bibfnamefont {I.~S.}\ \bibnamefont {Nefedov}}, \bibinfo {author} {\bibfnamefont {M.}~\bibnamefont {Silveirinha}}, \bibinfo {author} {\bibfnamefont {C.~R.}\ \bibnamefont {Simovski}},\ and\ \bibinfo {author} {\bibfnamefont {S.~A.}\ \bibnamefont {Tretyakov}},\ }\bibfield  {title} {\bibinfo {title} {Strong spatial dispersion in wire media in the very large wavelength limit},\ }\href@noop {} {\bibfield  {journal} {\bibinfo  {journal} {Physical Review B}\ }\textbf {\bibinfo {volume} {67}},\ \bibinfo {pages} {113103} (\bibinfo {year} {2003})}\BibitemShut {NoStop}%
\bibitem [{\citenamefont {Maslovski}\ \emph {et~al.}(2002)\citenamefont {Maslovski}, \citenamefont {Tretyakov},\ and\ \citenamefont {Belov}}]{maslovski2002wire}%
  \BibitemOpen
  \bibfield  {author} {\bibinfo {author} {\bibfnamefont {S.}~\bibnamefont {Maslovski}}, \bibinfo {author} {\bibfnamefont {S.}~\bibnamefont {Tretyakov}},\ and\ \bibinfo {author} {\bibfnamefont {P.}~\bibnamefont {Belov}},\ }\bibfield  {title} {\bibinfo {title} {Wire media with negative effective permittivity: a quasi-static model},\ }\href@noop {} {\bibfield  {journal} {\bibinfo  {journal} {Microwave and Optical Technology Letters}\ }\textbf {\bibinfo {volume} {35}},\ \bibinfo {pages} {47} (\bibinfo {year} {2002})}\BibitemShut {NoStop}%
\bibitem [{\citenamefont {Collin}(1990)}]{collin1990field}%
  \BibitemOpen
  \bibfield  {author} {\bibinfo {author} {\bibfnamefont {R.~E.}\ \bibnamefont {Collin}},\ }\href@noop {} {\emph {\bibinfo {title} {Field theory of guided waves}}},\ Vol.~\bibinfo {volume} {5}\ (\bibinfo  {publisher} {John Wiley \& Sons},\ \bibinfo {year} {1990})\BibitemShut {NoStop}%
\bibitem [{\citenamefont {Tretyakov}(2003)}]{tretyakov2003analytical}%
  \BibitemOpen
  \bibfield  {author} {\bibinfo {author} {\bibfnamefont {S.}~\bibnamefont {Tretyakov}},\ }\href@noop {} {\emph {\bibinfo {title} {Analytical modeling in applied electromagnetics}}}\ (\bibinfo  {publisher} {Artech House},\ \bibinfo {year} {2003})\BibitemShut {NoStop}%
\bibitem [{\citenamefont {Balanis}(2016)}]{balanis2016antenna}%
  \BibitemOpen
  \bibfield  {author} {\bibinfo {author} {\bibfnamefont {C.~A.}\ \bibnamefont {Balanis}},\ }\href@noop {} {\emph {\bibinfo {title} {Antenna theory: analysis and design}}}\ (\bibinfo  {publisher} {John wiley \& sons},\ \bibinfo {year} {2016})\BibitemShut {NoStop}%
\bibitem [{\citenamefont {Belov}\ and\ \citenamefont {Simovski}(2005)}]{belov2005homogenization}%
  \BibitemOpen
  \bibfield  {author} {\bibinfo {author} {\bibfnamefont {P.~A.}\ \bibnamefont {Belov}}\ and\ \bibinfo {author} {\bibfnamefont {C.~R.}\ \bibnamefont {Simovski}},\ }\bibfield  {title} {\bibinfo {title} {Homogenization of electromagnetic crystals formed by uniaxial resonant scatterers},\ }\href@noop {} {\bibfield  {journal} {\bibinfo  {journal} {Physical Review E}\ }\textbf {\bibinfo {volume} {72}},\ \bibinfo {pages} {026615} (\bibinfo {year} {2005})}\BibitemShut {NoStop}%
\bibitem [{\citenamefont {Maslovski}\ and\ \citenamefont {Silveirinha}(2009)}]{maslovski2009nonlocal}%
  \BibitemOpen
  \bibfield  {author} {\bibinfo {author} {\bibfnamefont {S.~I.}\ \bibnamefont {Maslovski}}\ and\ \bibinfo {author} {\bibfnamefont {M.~G.}\ \bibnamefont {Silveirinha}},\ }\bibfield  {title} {\bibinfo {title} {Nonlocal permittivity from a quasistatic model for a class of wire media},\ }\href@noop {} {\bibfield  {journal} {\bibinfo  {journal} {Physical Review B}\ }\textbf {\bibinfo {volume} {80}},\ \bibinfo {pages} {245101} (\bibinfo {year} {2009})}\BibitemShut {NoStop}%
\bibitem [{\citenamefont {Mariotti}\ and\ \citenamefont {Sankaran}(2010)}]{mariotti2010microplasmas}%
  \BibitemOpen
  \bibfield  {author} {\bibinfo {author} {\bibfnamefont {D.}~\bibnamefont {Mariotti}}\ and\ \bibinfo {author} {\bibfnamefont {R.~M.}\ \bibnamefont {Sankaran}},\ }\bibfield  {title} {\bibinfo {title} {Microplasmas for nanomaterials synthesis},\ }\href@noop {} {\bibfield  {journal} {\bibinfo  {journal} {Journal of Physics D: Applied Physics}\ }\textbf {\bibinfo {volume} {43}},\ \bibinfo {pages} {323001} (\bibinfo {year} {2010})}\BibitemShut {NoStop}%
\bibitem [{\citenamefont {Chen}(2002)}]{chen2002impedance}%
  \BibitemOpen
  \bibfield  {author} {\bibinfo {author} {\bibfnamefont {Z.}~\bibnamefont {Chen}},\ }\bibfield  {title} {\bibinfo {title} {Impedance matching for one atmosphere uniform glow discharge plasma (oaugdp) reactors},\ }\href@noop {} {\bibfield  {journal} {\bibinfo  {journal} {IEEE transactions on Plasma Science}\ }\textbf {\bibinfo {volume} {30}},\ \bibinfo {pages} {1922} (\bibinfo {year} {2002})}\BibitemShut {NoStop}%
\bibitem [{\citenamefont {Gautam}\ \emph {et~al.}(2021)\citenamefont {Gautam}, \citenamefont {Morra},\ and\ \citenamefont {Venkattraman}}]{gautam2021high}%
  \BibitemOpen
  \bibfield  {author} {\bibinfo {author} {\bibfnamefont {S.}~\bibnamefont {Gautam}}, \bibinfo {author} {\bibfnamefont {G.}~\bibnamefont {Morra}},\ and\ \bibinfo {author} {\bibfnamefont {A.}~\bibnamefont {Venkattraman}},\ }\bibfield  {title} {\bibinfo {title} {High frequency impedance characteristics of a tunable microplasma device},\ }\href@noop {} {\bibfield  {journal} {\bibinfo  {journal} {Journal of Applied Physics}\ }\textbf {\bibinfo {volume} {129}} (\bibinfo {year} {2021})}\BibitemShut {NoStop}%
\bibitem [{\citenamefont {Jacob}\ \emph {et~al.}(2006)\citenamefont {Jacob}, \citenamefont {Alekseyev},\ and\ \citenamefont {Narimanov}}]{Narimanov}%
  \BibitemOpen
  \bibfield  {author} {\bibinfo {author} {\bibfnamefont {Z.}~\bibnamefont {Jacob}}, \bibinfo {author} {\bibfnamefont {L.~V.}\ \bibnamefont {Alekseyev}},\ and\ \bibinfo {author} {\bibfnamefont {E.}~\bibnamefont {Narimanov}},\ }\bibfield  {title} {\bibinfo {title} {Optical hyperlens: Far-field imaging beyond the diffraction limit},\ }\href {https://doi.org/10.1364/OE.14.008247} {\bibfield  {journal} {\bibinfo  {journal} {Opt. Express}\ }\textbf {\bibinfo {volume} {14}},\ \bibinfo {pages} {8247} (\bibinfo {year} {2006})}\BibitemShut {NoStop}%
\bibitem [{\citenamefont {Sun}\ \emph {et~al.}(2015)\citenamefont {Sun}, \citenamefont {Shalaev},\ and\ \citenamefont {Litchinitser}}]{Shalaev}%
  \BibitemOpen
  \bibfield  {author} {\bibinfo {author} {\bibfnamefont {J.}~\bibnamefont {Sun}}, \bibinfo {author} {\bibfnamefont {M.~I.}\ \bibnamefont {Shalaev}},\ and\ \bibinfo {author} {\bibfnamefont {N.~M.}\ \bibnamefont {Litchinitser}},\ }\bibfield  {title} {\bibinfo {title} {Experimental demonstration of a non-resonant hyperlens in the visible spectral range},\ }\href@noop {} {\bibfield  {journal} {\bibinfo  {journal} {Nature Communications}\ }\textbf {\bibinfo {volume} {6}},\ \bibinfo {pages} {7201} (\bibinfo {year} {2015})}\BibitemShut {NoStop}%
\bibitem [{\citenamefont {Lu}\ and\ \citenamefont {Liu}(2012)}]{hyperreview}%
  \BibitemOpen
  \bibfield  {author} {\bibinfo {author} {\bibfnamefont {D.}~\bibnamefont {Lu}}\ and\ \bibinfo {author} {\bibfnamefont {Z.}~\bibnamefont {Liu}},\ }\bibfield  {title} {\bibinfo {title} {Hyperlenses and metalenses for far-field super-resolution imaging},\ }\href@noop {} {\bibfield  {journal} {\bibinfo  {journal} {Nature Communications}\ }\textbf {\bibinfo {volume} {3}},\ \bibinfo {pages} {1205} (\bibinfo {year} {2012})}\BibitemShut {NoStop}%
\bibitem [{\citenamefont {Zhao}\ \emph {et~al.}(2010)\citenamefont {Zhao}, \citenamefont {Palikaras}, \citenamefont {Belov}, \citenamefont {Dubrovka}, \citenamefont {Simovski}, \citenamefont {Hao},\ and\ \citenamefont {Parini}}]{zhao2010magnification}%
  \BibitemOpen
  \bibfield  {author} {\bibinfo {author} {\bibfnamefont {Y.}~\bibnamefont {Zhao}}, \bibinfo {author} {\bibfnamefont {G.}~\bibnamefont {Palikaras}}, \bibinfo {author} {\bibfnamefont {P.~A.}\ \bibnamefont {Belov}}, \bibinfo {author} {\bibfnamefont {R.~F.}\ \bibnamefont {Dubrovka}}, \bibinfo {author} {\bibfnamefont {C.~R.}\ \bibnamefont {Simovski}}, \bibinfo {author} {\bibfnamefont {Y.}~\bibnamefont {Hao}},\ and\ \bibinfo {author} {\bibfnamefont {C.~G.}\ \bibnamefont {Parini}},\ }\bibfield  {title} {\bibinfo {title} {Magnification of subwavelength field distributions using a tapered array of metallic wires with planar interfaces and an embedded dielectric phase compensator},\ }\href@noop {} {\bibfield  {journal} {\bibinfo  {journal} {New Journal of Physics}\ }\textbf {\bibinfo {volume} {12}},\ \bibinfo {pages} {103045} (\bibinfo {year} {2010})}\BibitemShut {NoStop}%
\bibitem [{\citenamefont {Brownless}\ \emph {et~al.}(2015)\citenamefont {Brownless}, \citenamefont {Sturmberg}, \citenamefont {Argyros}, \citenamefont {Kuhlmey},\ and\ \citenamefont {De~Sterke}}]{brownless2015guided}%
  \BibitemOpen
  \bibfield  {author} {\bibinfo {author} {\bibfnamefont {J.~S.}\ \bibnamefont {Brownless}}, \bibinfo {author} {\bibfnamefont {B.~C.}\ \bibnamefont {Sturmberg}}, \bibinfo {author} {\bibfnamefont {A.}~\bibnamefont {Argyros}}, \bibinfo {author} {\bibfnamefont {B.~T.}\ \bibnamefont {Kuhlmey}},\ and\ \bibinfo {author} {\bibfnamefont {C.~M.}\ \bibnamefont {De~Sterke}},\ }\bibfield  {title} {\bibinfo {title} {Guided modes of a wire medium slab: comparison of effective medium approaches with exact calculations},\ }\href@noop {} {\bibfield  {journal} {\bibinfo  {journal} {Physical Review B}\ }\textbf {\bibinfo {volume} {91}},\ \bibinfo {pages} {155427} (\bibinfo {year} {2015})}\BibitemShut {NoStop}%
\bibitem [{\citenamefont {Zhou}\ \emph {et~al.}(2010)\citenamefont {Zhou}, \citenamefont {Li}, \citenamefont {Yang}, \citenamefont {Peng}, \citenamefont {Su}, \citenamefont {Zhang}, \citenamefont {Li}, \citenamefont {Kim}, \citenamefont {Yu}, \citenamefont {Zhou} \emph {et~al.}}]{zhou2010plasmon}%
  \BibitemOpen
  \bibfield  {author} {\bibinfo {author} {\bibfnamefont {Z.-K.}\ \bibnamefont {Zhou}}, \bibinfo {author} {\bibfnamefont {M.}~\bibnamefont {Li}}, \bibinfo {author} {\bibfnamefont {Z.-J.}\ \bibnamefont {Yang}}, \bibinfo {author} {\bibfnamefont {X.-N.}\ \bibnamefont {Peng}}, \bibinfo {author} {\bibfnamefont {X.-R.}\ \bibnamefont {Su}}, \bibinfo {author} {\bibfnamefont {Z.-S.}\ \bibnamefont {Zhang}}, \bibinfo {author} {\bibfnamefont {J.-B.}\ \bibnamefont {Li}}, \bibinfo {author} {\bibfnamefont {N.-C.}\ \bibnamefont {Kim}}, \bibinfo {author} {\bibfnamefont {X.-F.}\ \bibnamefont {Yu}}, \bibinfo {author} {\bibfnamefont {L.}~\bibnamefont {Zhou}}, \emph {et~al.},\ }\bibfield  {title} {\bibinfo {title} {Plasmon-mediated radiative energy transfer across a silver nanowire array via resonant transmission and subwavelength imaging},\ }\href@noop {} {\bibfield  {journal} {\bibinfo  {journal} {ACS nano}\ }\textbf {\bibinfo {volume} {4}},\ \bibinfo {pages} {5003} (\bibinfo {year} {2010})}\BibitemShut {NoStop}%
\bibitem [{\citenamefont {Silveirinha}\ \emph {et~al.}(2008)\citenamefont {Silveirinha}, \citenamefont {Belov},\ and\ \citenamefont {Simovski}}]{silveirinha2008ultimate}%
  \BibitemOpen
  \bibfield  {author} {\bibinfo {author} {\bibfnamefont {M.~G.}\ \bibnamefont {Silveirinha}}, \bibinfo {author} {\bibfnamefont {P.~A.}\ \bibnamefont {Belov}},\ and\ \bibinfo {author} {\bibfnamefont {C.~R.}\ \bibnamefont {Simovski}},\ }\bibfield  {title} {\bibinfo {title} {Ultimate limit of resolution of subwavelength imaging devices formed by metallic rods},\ }\href@noop {} {\bibfield  {journal} {\bibinfo  {journal} {Optics letters}\ }\textbf {\bibinfo {volume} {33}},\ \bibinfo {pages} {1726} (\bibinfo {year} {2008})}\BibitemShut {NoStop}%
\bibitem [{\citenamefont {Aleshire}\ \emph {et~al.}(2020)\citenamefont {Aleshire}, \citenamefont {Pavlovetc}, \citenamefont {Collette}, \citenamefont {Kong}, \citenamefont {Rack}, \citenamefont {Zhang}, \citenamefont {Masiello}, \citenamefont {Camden}, \citenamefont {Hartland},\ and\ \citenamefont {Kuno}}]{aleshire2020far}%
  \BibitemOpen
  \bibfield  {author} {\bibinfo {author} {\bibfnamefont {K.}~\bibnamefont {Aleshire}}, \bibinfo {author} {\bibfnamefont {I.~M.}\ \bibnamefont {Pavlovetc}}, \bibinfo {author} {\bibfnamefont {R.}~\bibnamefont {Collette}}, \bibinfo {author} {\bibfnamefont {X.-T.}\ \bibnamefont {Kong}}, \bibinfo {author} {\bibfnamefont {P.~D.}\ \bibnamefont {Rack}}, \bibinfo {author} {\bibfnamefont {S.}~\bibnamefont {Zhang}}, \bibinfo {author} {\bibfnamefont {D.~J.}\ \bibnamefont {Masiello}}, \bibinfo {author} {\bibfnamefont {J.~P.}\ \bibnamefont {Camden}}, \bibinfo {author} {\bibfnamefont {G.~V.}\ \bibnamefont {Hartland}},\ and\ \bibinfo {author} {\bibfnamefont {M.}~\bibnamefont {Kuno}},\ }\bibfield  {title} {\bibinfo {title} {Far-field midinfrared superresolution imaging and spectroscopy of single high aspect ratio gold nanowires},\ }\href@noop {} {\bibfield  {journal} {\bibinfo  {journal} {Proceedings of the National Academy of Sciences}\ }\textbf {\bibinfo {volume} {117}},\ \bibinfo {pages} {2288} (\bibinfo {year}
  {2020})}\BibitemShut {NoStop}%
\end{thebibliography}%


\end{document}